\documentclass[10pt]{iopart}
\begin{document}
\title[Hamiltonian dynamics of extended objects]
{Hamiltonian dynamics of extended objects}
\author{R  Capovilla\dag ,
J Guven\ddag\S  and E Rojas\P}
\address{\dag\
Departamento de F\'{\i}sica,
Centro de Investigaci\'on
y de Estudios Avanzados del IPN,
Apdo Postal 14-740, 07000 M\'exico,
D. F.,
MEXICO}
\address{\ddag\ School of Theoretical Physics,
Dublin Institute for Advanced Studies,
10 Burlington Road, Dublin 4, IRELAND}
\address{\S\  
Instituto de Ciencias Nucleares,
Universidad Nacional Aut\'onoma de M\'exico,
Apdo. Postal 70-543, 04510 M\'exico, DF, MEXICO}
\address{\P\ Facultad de F\'{\i}sica e Inteligencia 
Artificial, 
Universidad Veracruzana, 91000 Xalapa, Veracruz, MEXICO}

\begin{abstract}
We consider a relativistic extended
object described by a reparametrization invariant local action
that depends on the extrinsic curvature of the worldvolume
swept out by the object as it evolves. We 
provide a Hamiltonian
formulation of the dynamics of such higher derivative models which is
motivated by the ADM formulation of general relativity. 
The canonical momenta are identified 
by looking at boundary behavior under small 
deformations of the  action; the relationship 
between the momentum conjugate to the embedding functions and 
the conserved momentum density is established. 
The canonical Hamiltonian is constructed explicitly;
the constraints on the phase space, both primary and secondary, 
are identified and the role they play in the theory
described.  The multipliers implementing the primary constraints 
are identified in terms of the ADM lapse and shift variables 
and Hamilton's equations 
shown to be consistent with the Euler-Lagrange equations.
\end{abstract}

\pacs{87.16.Dg, 46.70.Hg}




\section{Introduction}

Relativistic extended objects, such as strings and membranes,
can be viewed either as fundamental building blocks of field theories, 
or as providing, in their own right,  
field theories with an intrinsically geometric content. 
The geometry of interest is the trajectory,  or worldvolume,
swept out by the relativistic object in its evolution in a fixed
background spacetime. The 
dynamics of this object is described by a local action constructed using the
geometric scalars that characterize the worldvolume geometry.
The dual requirements of worldvolume reparametrization invariance and
invariance under the ambient spacetime symmetries 
place severe limits  on the possible scalars that can be constructed.
At lowest order, there is the Dirac-Nambu-Goto
[DNG]
action, proportional to the area of the worldvolume \cite{DNG1,DNG2,DNG3}.
This action is of first order in derivatives of the field variables
-- the embedding functions for the worldvolume.
There are many contexts, however, where it appears that
the DNG action is
not adequate, and it is natural to consider models that
depend on higher derivatives of these variables. Examples
include the addition of a rigidity term in 
an effective action for stringy QCD \cite{Polyakov,Kleinert}, the 
systematic approximations that arise in   
expansions in the thickness of topological defects 
\cite{Corrections1,Corrections2,Corrections3,Corrections4},  
or actions that appear in braneworld scenarios 
\cite{Braneworld}. 
The dependence
on derivatives of the embedding functions higher than first
can appear only through the extrinsic and intrinsic curvatures
of the worldvolume and their covariant 
derivatives. It is natural to cast the model in terms of these 
curvatures. Models of this kind also show up in the more formal 
context of non-linear
physics in the description of integrable surfaces
\cite{Integrable}.
Their Euclidean counterparts are also of interest as  
fluctuating surfaces \cite{fluctua1,fluctua2}, and in soft condensed 
matter physics, {\it e.g.} in the theoretical description of
lipid vesicles \cite{vesicles1,vesicles2}.

In this paper, we examine the Hamiltonian formulation 
of the dynamics of relativistic extended objects described
by an action that depends on the extrinsic  
curvature of its worldvolume.  A dependence of the action on the
worldvolume intrinsic curvature can
always be reduced to one depending on the appropriate combinations
of the extrinsic curvature using the Gauss-Codazzi equations. Of course,
this topic has been explored before for particular models (see {\it e.g.}
\cite{Nesterenko1,Nesterenko2}).
However, 
a thorough canonical analysis is still missing.
Previous approaches have typically exploited auxiliary 
variables, artificially reducing the higher derivative
theory to an action that depends at most on first derivatives of 
an augmented set of field variables. This strategy has been valuable
in the perturbative evaluation of functional integrals
for geometric models \cite{Polyakovbook}. At the Hamiltonian level, however, it
has the drawback of introducing artificial secondary constraints
that obscure the canonical structure. 
It is all too easy to lose sight of the geometrical object one started out with.
Another  approach has been focused on the general structure of diffeomorphism
invariant actions \cite{Hoppe1,Hoppe2,Hoppe3,Hoppe4}, but
it does not address the peculiar issues of higher 
derivative models. Covariant symplectic formulations of geometric
models for extended objects have been 
examined in \cite{Cartas,Carter}.

Our approach is to grapple directly with the 
higher derivative nature of geometric models that depend 
on the extrinsic curvature. In the Hamiltonian formulation, 
the focus is shifted from the worldvolume geometry 
to the time-dependent geometry of the extended object. 
This approach will 
complement the Lagrangian approach taken in \cite{Defo} (an alternative
treatment of the dynamics of extended objects that emphasizes
covariance with respect to the ambient spacetime has been
advocated by Carter \cite{Carterbrane}).

It is first necessary to  develop an appropriate geometric language to
accommodate the arbitrariness in the foliation of the worldvolume by
the object as it evolves.
To do this, we are guided by the ADM formulation of canonical general 
relativity \cite{ADM},
where spacetime is constructed by the evolution of three-dimensional
spatial slices. 
The ambiguity inherent in the description of the 
evolution is captured by the normal and tangential projections of the velocity vector:
the lapse and shift variables. 

We start by considering in Sect. 2 the Hamiltonian formulation of a DNG
model. In order to minimize formalism, initially we restrict our attention
to a relativistic object of dimension $d$ 
propagating in a flat Minkowski spacetime of 
dimension $d+2$; the worldvolume of dimension $d+1$ is a hypersurface
with a single normal. The generalization to higher co-dimension is
straightforward. The case of a relativistic
string with $d=1$ is quite special; since it is one-dimensional its intrinsic
geometry is trivial. The DNG
action is a useful toy model in that it 
provides a benchmark for various features of the formalism which we 
will develop; in this case the relationship between 
the geometry of the worldvolume and that of the object is particularly
simple.
The phase space is given by the embedding functions of the object
at `fixed time' and its conjugate momentum, proportional to the
timelike unit normal from the object onto the worldvolume; an immediate
generalization of the momentum for a massive relativistic particle.
Because of reparametrization invariance, the Hamiltonian vanishes;
it is a linear combination of constraints. The constraints are first class
and they  are the generators of worldvolume reparametrizations.
Next, in Sect. 3, we illustrate the main features of the general Hamiltonian 
formulation for an extended object with a dependence on second derivatives of
the field variables.  At face value this is a straightforward generalization
of the classical Ostrogradski Hamiltonian 
treatment of higher derivative theories
to extended objects \cite{Ostro}: in short partial derivatives get promoted 
to functional derivatives; the phase space  
is extended to include the 
velocity of the embedding functions for the extended object and its conjugate
momentum; the Hamiltonian is given by a Legendre transformation
with respect to both velocities and accelerations.
There are subtleties, however, even at this level. We will show how the 
introduction of the ADM variables facilitates
the identification of the momenta 
as well as the implementation of the Lagrange transformation in the construction
of the Hamiltonian.  Of course, the principal subtlety is the 
identitication of the constraints on the canonical variables associated with
the underlying reparametrization invariance and understanding the role
they play in the theory. 
For clarity, we have opted not to treat 
these issues in full generality, addressing them only for 
a class of `non-degenerate' models
in which the momentum conjugate to the velocity may be inverted for
the acceleration, and in particular, 
for the model quadratic in the mean extrinsic curvature.
A model in which
the dependence on the acceleration of the action is only linear, such as 
a model linear in the mean extrinsic curvature, behaves very differently; 
we will consider it elsewhere. 
In order to  identify the extended phase space for these models,
in Sect. 4 we consider the first variation of the action in a geometrical way.
By examining the boundary behaviour of
the varied action, we identify both the momenta conjugate to
the embedding functions and those
conjugate to their
velocities. The latter is always normal to the 
worldvolume. The former  is a sum of a bulk (curvature dependent) term 
and a total spatial divergence.  
In Sect. 5, we compare the structure of the canonical momenta
with the conserved linear and angular momentum 
associated with Poincar\'e symmetry \cite{Noether}. 
In Sect. 6
we focus on the derivation of the canonical Hamiltonian
for a model quadratic in the mean extrinsic curvature, and we identify the
primary constraints. Together with the canonical Hamiltonian these
constraints generate the dynamics. 
There are secondary constraints; 
one of these is the vanishing of the canonical Hamiltonian.
The secondary constraints are the generators of worldvolume
diffeomorphisms. 
There are no tertiary constraints, and the
constraints are in involution; they are first class.   
In Sect. 7 we check explicitly that Hamilton's equations 
for the model quadratic in the mean extrinsic curvature
are completely equivalent to the vanishing of its Euler-Lagrange derivative. 
We obtain explicit expressions for the Lagrange multipliers
that enforce the constraints. Unlike the DNG model, they are related in a
non-trivial way to the ADM lapse and shift variables,
a feature which complicates the comparison between 
the Lagrangian and Hamiltonian formulations of the theory.
In Sect. 8 we lift the restriction to co-dimension one for the
worldvolume, and we show that the Hamiltonian formulation is
essentially unchanged. We conclude in Sect. 9 with some remarks.
We have collected in Appendix A some useful formulas 
describing the geometries
of the worldvolume and of the extended object. In Appendix B, we give
the full Poisson algebra of the constraints for the DNG model and for the
model quadratic in the mean extrinsic curvature.

In this paper, we will assume that the extended object has no spatial 
boundary.

\section{Dirac-Nambu-Goto}

We consider a relativistic extended object $\Sigma$, of dimension $d$,
in a fixed background Minkowski spacetime of dimension $d+2$. As 
$\Sigma$ 
evolves, it describes its trajectory or worldvolume  $m$, of dimension
$d+1$. We 
will be concerned with both the geometries of $\Sigma$ and $m$; some
notational clutter is unavoidable. The worldvolume $m$ is given by the
timelike embedding 
\begin{equation}
x^\mu = X^\mu (\xi^a ) = X^\mu (t, u^A )\,,
\end{equation}
where  $ x^\mu$ are local coordinates for $M^{d+2}$ ($\mu,\nu,\dots = 
0,1,\cdots, d+1$),  $\xi^a$ local coordinates
for the worldvolume $m$ ($a,b, \dots = 0,1,2, \cdots, d$), and $X^\mu$
are 
the embedding 
functions for $m$. In the second equality, we split the local coordinates
for $m$ in an arbitrary evolution parameter $t$ and the coordinates $u^A$ 
for $\Sigma$ at fixed values of $t$
($A,B,\dots = 1,2,\cdots, d)$. 
In spirit and in many technical
details, this split is similar to the ADM approach to canonical general
relativity (see {\it e.g.} \cite{ADM}).  Note that the object $\Sigma$ 
can be 
presented in two ways: as the spacelike embedding in spacetime $ x^\mu = 
X^\mu (t=0, u^A)$, or as the spacelike embedding in the worldvolume $m$,
$\xi^a = X^a (u^A)$, related by composition.
Recall that we assume that the extended object $\Sigma$ has no spatial boundary.
(We have collected some useful formulae describing the geometries
of $m$ and $\Sigma$ in Appendix A.)

In this
section, we focus on the Hamiltonian formulation of an object whose
dynamics is determined by the DNG 
action, proportional to the area of the worldvolume swept out by 
the object in its 
evolution (see {\it e.g.} \cite{Dolan,Smolin,ADMDNG}). 
The basic field variables are the embedding functions 
for the worldvolume, and this action depends only on first derivatives
of the field variables. The DNG action is the generalization to
extended objects of the action for a relativistic particle
in its geometrical square root form. 
The action is
\begin{equation}
S [ X] =-\mu \; \int_m d^{d+1}\xi \; \sqrt{-g} = - \mu \int_m 
\; \sqrt{-g}\,.
\label{eq:DNG1}
\end{equation}
Henceforth, we will absorb the differentials $d^{d+1} \xi$ in the integral 
sign. The constant $\mu$ is the tension. $g$ denotes the 
determinant of the induced metric on $m$, 
\begin{equation}
g_{ab} = \eta_{\mu\nu} X^\mu_a   X^\nu_b = X_a \cdot  X_b\,,
\label{eq:wmetric}
\end{equation}
where $X_a = \partial X / \partial \xi^a$ are the 
$d+1$ tangent vectors to $m$, and $\eta_{\mu\nu}$ is the Minkowski 
metric with one minus sign. We will denote by $\nabla_a$ the 
worldvolume covariant derivative compatible with $g_{ab}$.
Latin indices are lowered 
and raised with $g_{ab}$ and its inverse $g^{ab}$, respectively.

The first step towards a Hamiltonian formulation is to
consider the first variation of the action under an infinitesimal
deformation of the embedding functions $X \to X + 
\delta X$. We readily obtain
\begin{eqnarray}
\delta S [X] &=&
- \mu \int_m \; \sqrt{-g} 
g^{ab}  X_a \cdot \partial_b 
\delta  X
\nonumber \\
&=&  \mu \int_m \; 
\partial_b [ \sqrt{-g} g^{ab} X_a ] 
\cdot 
 \delta  X - 
 \mu \int_m  \; \partial_b [ 
\sqrt{-g} g^{ab} X_a \cdot 
\delta X]\,.
\label{eq:vardng1}
\end{eqnarray}
We have used the fact that $\delta X_a = \partial_a \delta X$,
and we have integrated by parts to obtain the second line.
We note that, using Stokes' theorem, for an arbitrary worldvolume vector $V^a$, 
we have
\begin{equation}
\int_m \; \partial_a \; ( \sqrt{-g} \; V^a )
= - \int_\Sigma d^d u \; \sqrt{h} \; \eta_a V^a
= - \int_\Sigma \; \sqrt{h} \; \eta_a V^a
 \,.
\label{eq:stokes}
\end{equation}
Here $\eta^a$ is the timelike unit normal to $\Sigma$ onto $m$, 
$g_{ab} \eta^a \eta^b = -1$,
and the minus sign on the right hand side  of (\ref{eq:stokes}) comes 
about because of the direction of the normal $\eta$;
$h$ denotes
the determinant of the metric induced on $\Sigma$, 
\begin{equation}
h_{AB} = X_A \cdot X_B = g_{ab} X^a_A X^b_B \,,
\end{equation}
with either $X_A = \partial X / \partial u^A$ or
$X^a_A = \partial X^a / \partial u^A $ the $d$ tangent vectors to 
$\Sigma$. Capital latin indices are lowered and raised with $h_{AB}$
and its inverse $h^{AB}$, respectively. We will denote with
${\cal D}_A = X^a_A \nabla_a$ the spatial covariant derivative
on $\Sigma$, compatible with $h_{AB}$.
Note that $g_{ab} \eta^a X^b_A = 0$. The worldvolume
vectors $\{ \eta^a , X^a_A \}$ form a basis for $m$ adapted to
$\Sigma$. They satisfy the completeness relation
\begin{equation} 
g^{ab} = - \eta^a \eta^b + h^{AB} X^a_A X^b_B\,. 
\label{eq:compli}
\end{equation}

One convenient way to obtain the  canonical momentum $p$ conjugate to the
embedding functions $X$ is to  recall that
the first variation of the action can be written, for a first 
order theory, as
\begin{equation}
\delta S [ X ] =  \int_m   \; \sqrt{-g} \; E
\cdot  \delta X +
\int_\Sigma  \; 
p \cdot
\delta X\,,
\label{eq:varp1}
\end{equation}
where $E$ denotes the Euler-Lagrange derivative of $S[X]$.
From (\ref{eq:vardng1}) and (\ref{eq:stokes}),  we can cast the first
variation of the action in the form (\ref{eq:varp1}) with
\begin{equation}
\delta S [ X  ] =
 \mu \int_m  \; 
\partial_b [ \sqrt{-g} g^{ab} X_a ] 
\cdot 
\delta  X + 
 \mu \int_\Sigma   \; \sqrt{h} \; \eta_b \; g^{ab} X_a \cdot 
\delta X\,,
\label{eq:vardng2}
\end{equation}
so that the canonical momenta
can be read off 
from the boundary 
term as
\begin{equation}
p =   \mu \; \sqrt{h} \; \eta^a X_a 
=  \mu \; \sqrt{h} \; {\bf \eta}\,.
\label{eq:mom}
\end{equation}
It is proportional to the timelike unit normal to $\Sigma$ into the 
worldsheet. We write
$ \eta = \eta^a 
X_a$; $p$ is a spatial density of weight one.
Note that in the limit of a pointlike object we have that 
$h=1$, $\mu$ is the mass, and $\eta$ the relativistic velocity.

We have thus identified the phase space for the DNG model: the embedding
functions for the relativistic object $\Sigma$ are the canonical coordinates,
and their conjugate canonical momenta are proportional to the densitized
timelike unit normal to $\Sigma$ into the worldvolume. 

From (\ref{eq:vardng2}),  we can also identify  the 
Euler-Lagrange derivative of the DNG action as
\begin{equation}
E =  - \mu \; {1 \over \sqrt{-g}} \; \partial_b [ \sqrt{-g} g^{ab} X_a ] 
= - \mu \; \nabla^2 X = - \mu \; K \; n\,,
\label{eq:eldng}
\end{equation}
where $\nabla^2 = g^{ab} \nabla_a \nabla_b$ denotes the d'Alembert
operator, and $K$ is the mean 
extrinsic curvature, defined by $K = g^{ab} K_{ab}$ with 
\begin{equation}
K_{ab} =
- n \cdot \partial_a X_b
\end{equation}
the extrinsic curvature
of $m$, and $n$ the spacelike unit normal to $m$, defined implicitly
by $n \cdot X_a = 0$, $  n \cdot n = 1$. 
It follows that the equations of motion for a DNG object can be expressed 
succintly as the vanishing of the mean extrinsic curvature
\begin{equation}
- \mu \; K  = 0\,.
\label{eq:eomdng}
\end{equation}

We obtain the canonical Hamiltonian via a Legendre transformation
with respect to the velocity of the embedding functions for $\Sigma$, 
\begin{equation}
H_c  [ X, p ]
= \int_\Sigma \;  p \cdot \dot{X} - 
L [ X, \dot{X} ]\,.
\end{equation}
The velocity $\dot{X} = \partial_t X = \partial_0 X$ 
is a vector tangent to the worldvolume $m$ (note that we were able to 
identify $p$ without any explicit reference to $\dot {X}$). 
It  can be expanded in 
components with respect to the basis of tangent vectors to the worldvolume, 
$\{ \eta , 
X_A \}$ adapted to $\Sigma$ as
\begin{equation}
\dot{X} = N \;\eta + N^A \; X_A \,.
\label{eq:velo}
\end{equation}
Using a language borrowed
from canonical general relativity, the projections
of $\dot{X}$ onto this basis are 
the lapse function $N$ and the shift vector $N^A$.  
In this model, neither $N$ nor $N^A$ is a canonical variable;
in the higher derivative theory we will see that they both are.

With respect to the coordinate basis $\{ \dot{X}, X_A\}$ adapted to the
evolution, the worldvolume 
metric (\ref{eq:wmetric}) takes its ADM form
\begin{equation}
g_{ab}= \left(
\begin{array}{cc}
-N^2 + N^A N^B h_{AB} & h_{AB}N^B \\
h_{AB}N^B & h_{AB}
\end{array}
\right)\,,
\label{eq:metric}
\end{equation}
and its  determinant is given by
\begin{equation}
g = - N^2 h\,,
\label{eq:det}
\end{equation}
in terms of the lapse function $N$ and the determinant $h$ of the spatial
metric 
$h_{AB}$. We will also need the ADM form of the inverse metric, 
\begin{equation}
g^{ab}= {1 \over N^2} \left(
\begin{array}{cc}
-1 & N^A \\
N^A  & N^2 h^{AB} - 
N^{A}N^{B}
\end{array}
\right)\,.
\label{eq:inverse-metric}
\end{equation}

The DNG Lagrangian functional is 
\begin{equation}
S = \int dt \; L [ X, \dot{X} ]\,,
\end{equation}
where
\begin{equation}
L [ X, \dot{X} ] = - \mu \int_\Sigma  \; N \; \sqrt{h}\,,
\end{equation}
and  we have used (\ref{eq:det}).
Note that the canonical momenta can also be obtained directly by the
functional derivative
$p = \delta  L / \delta \dot{X} = \mu \sqrt{h} \eta$,
since $N = - \eta \cdot \dot{X}$, in agreement with (\ref{eq:mom}). 
However, this direct derivation  becomes difficult to carry out for higher 
derivative actions.

Using (\ref{eq:mom}),  the canonical Hamiltonian is given by
\begin{equation}
H_c [ X, p ]
= \int_\Sigma \; [ \mu \; \sqrt{h} \; \eta \cdot \dot{X} 
+ \mu N \sqrt{h} ] = 0\,.
\end{equation}
It vanishes;  this was to be 
expected because of the 
reparametrization invariance of the DNG action. 
How this comes about is rather more subtle in the 
higher derivative theory.
According to the standard 
Dirac-Bergmann theory of constrained systems, the 
Hamiltonian is a linear combination of  constraints, 
which are easily identified from the
definition of the momenta (\ref{eq:mom})
as
\begin{eqnarray} 
C_A &=& p \cdot X_A =0\,,
\label{eq:CA}
\\
C &=& p^2 + \mu^2 \; h = 0\,.
\label{eq:C}
\end{eqnarray} 
Note that $C_A$ is a spatial scalar density of weight one, whereas
$C$ is a spatial scalar density of weight two. There are $d+1$ constraints.
It is easy to check that they are in involution, so they
are first class constraints (see Appendix B). They are the generators of
worldvolume reparametrizations; $C_A$ generates diffeomorphisms
tangential to $\Sigma$, whereas $C$ generates diffeomorphisms
out of $\Sigma$ onto the worldvolume $m$.
For a relativistic string, 
$h$ provides a quadratic potential in the constraint (\ref{eq:C});
for a relativistic membrane $h$ provides a quartic potential.
The number of physical degrees of freedom is given by $1/2 [$ 
(total number of canonical variables) $- 2 \times$ (number of first
class constraints)]. We obtain that there is a single physical degree
of freedom, which corresponds to motions along the single normal to
the worldvolume. For higher co-dimension there will be a physical
degree of freedom along each normal.

It follows that the  Hamiltonian that generates the dynamics is 
\begin{equation}
H [ X, p] =
\int_\Sigma 
\; ( \lambda \; C + \lambda^A \; C_A )\,,
\label{eq:hamdng}
\end{equation}
where $\lambda, \lambda^A$ are arbitrary Lagrange multipliers that
enforce the constraints. Note that $\lambda$ must be a 
spatial density of weight minus one for the integral over $\Sigma$
to be well-defined.

The Poisson brackets are, in terms of two arbitrary phase space functions
$f,g$,
\begin{equation}
\{ f , g \} = \int_\Sigma \left(
{\delta f \over \delta X } \cdot
{\delta g \over \delta p } -
{\delta g \over \delta X } \cdot
{\delta f \over \delta p } \right)\,.
\label{eq:poisson}
\end{equation}

Let us check explicitly that the Hamilton equations that follow 
from (\ref{eq:hamdng}) reproduce the equations of motion (\ref{eq:eomdng}). 
This exercise is useful to illustrate
various features of the Hamilton equations for an extended object.
The first Hamilton equation is
\begin{equation}
\dot{X} = {\delta H \over \delta p } =   2\lambda  \; p + 
\lambda^A \; X_A\,,
\label{eq:h1}
\end{equation}
and, as in its non-relativistic counterpart, it serves to express
the momenta in terms of the velocities.  Using  the 
constraints 
(\ref{eq:CA}), (\ref{eq:C}), the Lagrange multipliers are identified as
\begin{equation}
\lambda = {N \over 2\mu\sqrt{h}}\,, \quad \quad \quad
\lambda^A=N^A\,.
\label{eq:lagra}
\end{equation}
As we will see, this simple identification does not hold 
in higher order theories.  
With this identification, 
the Hamilton equation (\ref{eq:h1}) reproduces the form of
the canonical momenta (\ref{eq:mom}).
The second Hamilton equation is
\begin{eqnarray}
\dot{p} &=& - {\delta H \over \delta X } 
=
- \int_\Sigma 
\; \left( \lambda \mu^2 h h^{AB} {\delta h_{AB} \over 
\delta X}  + \lambda^A \; p \cdot {\delta X_A \over
\delta X} \right) \,,
\nonumber \\ 
 &=&  
\partial_A \left( 2 \mu^2  \lambda  h h^{AB}
X_B + N^A  p \right)
\,,
\end{eqnarray}
where we have integrated by parts, disregarded 
a total spatial divergence, and used
\begin{eqnarray}
\delta h_{AB} &=& 2 X_{(A} \cdot \partial_{B)} \delta X\,, \\
\delta h &=& 2 h h^{AB} X_A \cdot \partial_B \delta X\,.
\end{eqnarray} 
Use of the form 
(\ref{eq:lagra}) for the Lagrange multipliers
gives the equations of motion in canonical form as
\begin{equation}
 \dot{p}
= \partial_A \left(  \mu N \sqrt{h}
h^{AB}  X_{B} + N^A \, p \right)\,.
\label{eq:h2dng}
\end{equation}
The time derivative of the canonical momenta is given by
a spatial divergence. The total momentum is conserved.
This feature is common to any
Poincar\'e invariant action. The reason is simple:
because of Poincar\'e invariance the dependence on the embedding 
functions $X$
of the Hamiltonian can appear only through its spatial derivatives. 
Note that the right hand side
of (\ref{eq:h2dng}) is the divergence of a spatial density of
weight one; as such, it is independent of the spatial 
affine connection.

This canonical expression for the equations of motion should be compared 
with the worldvolume covariant equations of motion (\ref{eq:eomdng}). To see 
that 
they coincide, we decompose  the DNG Euler-Lagrange 
derivative appearing in (\ref{eq:eldng}) with respect to the basis
$\{ \dot{X} , X_A \}$, adapted to the evolution of $\Sigma$. We have, 
using (\ref{eq:det}), (\ref{eq:inverse-metric}) for
$\sqrt{-g}$ and $g^{ab}$, or, alternatively, 
\begin{eqnarray}
g^{0b} &=& - {1 \over N} \eta^b\,,
\label{eq:g0b}\\
g^{Ab} &=& h^{AB} X^b_B + {N^A \over N} \eta^b\,,
\label{eq:gAb}
\end{eqnarray}
that
\begin{eqnarray}
\sqrt{-g} E &=& - \mu \partial_0 \; ( N \sqrt{h} g^{0b} X_b )
- \mu \partial_A \; ( N \sqrt{h} g^{Ab} X_b )
\nonumber \\
&=&
\mu \partial_0 \; ( \sqrt{h} \eta )
- \mu \partial_A ( \sqrt{h}  
N^A \eta + N \sqrt{h} h^{AB} X_B )\,.
\end{eqnarray}
Now we need only to recognize that $ \mu \sqrt{h} \eta = p$, 
to obtain
\begin{equation}
\sqrt{-g} E = \dot{p} - \partial_A \left(  \mu N \sqrt{h}
h^{AB}  X_{B} + N^A \, p \right)\,,
\end{equation}
whose vanishing coincides with the canonical form of the 
equations of motion (\ref{eq:h2dng}).

\section{Higher derivatives models}

We turn now to geometric models for the dynamics of relativistic extended
objects that depend, in addition to the velocity as in the DNG model, also
on the acceleration of the embedding functions, $\ddot{X}$. We
consider a local action of the form
\begin{equation}
S[X] =  \int dt 
\; L [ X, \dot{X} , \ddot{X} ]\,,
\label{eq:scov}
\end{equation}
where $ L [ X, \dot{X} , \ddot{X} ] $ is the Lagrangian functional.
We are dealing with a higher derivatives theory. The
Hamiltonian formulation of  such theories \cite{Ostro} involves the 
extension of the phase space to $\{p, X; P , \dot{X} \}$, 
where $P$ and $p$
denote the momenta conjugate to $\dot{X}$ and $X$, respectively.
They can be obtained directly from the functional derivatives
of the Lagrangian functional
\begin{eqnarray}
P &=& 
{\delta L \over \delta \ddot{X}}\,,
\label{eq:bigp}
\\
p &=&   {\delta L \over \delta \dot{X}} -
\partial_0 \left({\delta L \over \delta \ddot{X}}\right)\,.
\label{eq:smallp}
\end{eqnarray}
Alternatively, the momenta can be read off from the boundary
term in the first order variation of the action, that, for
a theory that depends at most on second derivatives of the
field variables, can always be written in the form
\begin{equation}
\delta S [ X ] =  \int_m   \; \sqrt{-g} \; E
\cdot  \delta X +
\int_\Sigma    \; 
( P \cdot \delta \dot{X} + p \cdot
\delta X )\,,
\label{eq:varp}
\end{equation}
where $E$ denotes the Euler-Lagrange derivative. This is the higher 
derivative generalization of (\ref{eq:varp1}).

The canonical Hamiltonian is given by the Legendre transformation
with respect to both $\ddot{X}$ and $\dot{X}$ as
\begin{equation}
H_c [X,p;\dot{X}, P ] 
= \int_\Sigma  \left( P \cdot \ddot{X} + p \cdot \dot{X} \right)  - 
L [ X, \dot{X} , \ddot{X} ]\,.
\label{eq:hc1}
\end{equation}
In this expression, it is understood that the acceleration $\ddot{X}$
is expressed in terms of the phase space variables $P, \dot{X}, X$, just as
in a first order theory one expresses $\dot{X}$ in terms of $p$ and $X$.
It should be emphasized that $p$ is left alone. The resulting expression 
for the canonical Hamiltonian is
\begin{equation}
H_c [X,p;\dot{X}, P ] 
= \int_\Sigma ( p \cdot \dot{X}   + V [ X, \dot{X} , P])\,,
\label{eq:hcpot}
\end{equation}
where the form of the `potential' $V [X, \dot{X}, P ]$ will depend on the 
specific model under consideration.

The presence of symmetries manifests itself in the
canonical formalism by the appearance of constraints.
If there are primary constraints, they are identified by the null 
eigenvectors of the Hessian
\begin{equation}
{\cal H}_{\mu\nu} = {\delta^2 L [ X , \dot{X} , \ddot{X} ] \over 
\delta \ddot{X}^\mu \delta \ddot{X}^\nu }\,,
\label{eq:hess}
\end{equation}
and, if present, must be added to the canonical Hamiltonian (\ref{eq:hc1}).
This produces an extended Hamiltonian which generates the dynamics 
via the Hamilton equations. Rather than addressing this issue in full 
generality
here, we prefer to address it within a specific model below in Sect. 6.

The Poisson brackets (\ref{eq:poisson})  generalize to 
\begin{equation}
\{ f , g \} = \int_\Sigma \left[ {\delta f \over \delta X } \cdot
{\delta g \over \delta p } + 
{\delta f \over \delta \dot{X} } \cdot
{\delta g \over \delta P } - (f \leftrightarrow g ) \right]\,.
\end{equation}

Let us consider briefly the structure of the Hamilton equations.
The first one is vacuous, since the only dependence of the
Hamiltonian on $p$ is through the $p \cdot \dot{X}$ term,
$\dot{X} = \{ X , H \} = \delta H / \delta p  = \dot{X}\,.$
The second provides the definition of the momenta $P$
conjugate to $\dot{X}$, 
$\ddot{X} = \delta H / \delta P  \,,$
and it reproduces (\ref{eq:bigp}). Moreover, it identifies the
explicit form of the Lagrange multipliers that enforce the primary 
constraints, if present.  
The third Hamilton equation provides the definition of the momenta
$p$ conjugate to $X$,$
\dot{P} = - \delta H / \delta \dot{X}$, 
and it reproduces (\ref{eq:smallp}). Finally, the fourth equation gives the
equations of motion in
Hamiltonian form, once the expressions for $p,P$ are used:
$\dot{p} = - \delta H / \delta X $.

For a relativistic extended object, the requirement of reparametrization
invariance implies that the dependence on second derivatives of
the embedding functions can appear only through the extrinsic
curvature of the worldvolume $m$. Therefore, the explicitly covariant form 
of the action (\ref{eq:scov}) can be written as
\begin{equation}
S [ X ] = \int_m \sqrt{-g} \; L (g^{ab} , K_{ab} )\,,
\label{eq:haction}
\end{equation}
where $L (g^{ab} , K_{ab}) $ is a scalar constructed from the
inverse induced metric and the extrinsic curvature tensor
$K_{ab} = - n \cdot \partial_a X_b\,.$ 
A possible dependence on the intrinsic curvature of the
worldvolume ${\cal R}_{abcd}$ can always be expressed in terms of
the extrinsic curvature via the Gauss-Codazzi equations for the
worldvolume $m$,
\begin{equation}
{\cal R}_{abcd} = K_{ac} K_{bd} - K_{ad} K_{bc}\,.
\end{equation}

An important example of an action of the form (\ref{eq:haction})
is provided by the rigidity model, quadratic in the mean
extrinsic curvature,
\begin{equation}
S[X] =  {1 \over 2} \int_m  \sqrt{-g}\;
 K^2 \,.
\label{eq:haction2}
\end{equation}
For a relativistic string, the addition of this term dependent
on the extrinsic geometry to the DNG action has been suggested
by Polyakov \cite{Polyakov}, and independently by Kleinert 
\cite{Kleinert}, in an effective description of QCD. Aspects of
its
Hamiltonian formulation have been considered by Nesterenko
and Suan Han \cite{Nesterenko1}.

This quadratic model is representative of the generic case
at least as long as the rank of the Hessian (\ref{eq:hess})
does not change. What will change is the specific form of
the potential $V [ X , \dot{X}, P ]$ in the canonical 
Hamiltonian (\ref{eq:hcpot}).

One important exception is the case when the dependence
on the acceleration in (\ref{eq:haction}) is only linear.
An example is the Tolman model, valid only for a hypersurface,
linear in the mean extrinsic curvature \cite{Tolman}
\begin{equation}
S [X] = \int_m \sqrt{-g} \; K\,.
\end{equation} 
Another is the Einstein-Hilbert action for a relativistic
extended object \cite{Regge,Deser,Davidson}
\begin{equation}
S [X] = \int_m \sqrt{-g} \; {\cal R}\,,
\end{equation} 
which, for a relativistic string, by the Gauss-Bonnet theorem, is a
topological invariant.
The canonical analysis of these cases is qualitatively different. 
We will consider them elsewhere
\cite{CGREH}. The problem is
that it is not possible to carry out the Legendre transformation,
{\it i.e.} invert (\ref{eq:bigp}) to express $\ddot{X}$ in terms
of the extended phase space variables.

\section{First variation and momenta}

In this section, we use the first variation of the action to identify 
both the equations of motion and the 
canonical 
momenta conjugate to $\{ X, \dot{X} \}$ for the extrinsic curvature 
dependent  action (\ref{eq:haction}).
For the latter purpose, one  approach would be to carry out  directly the 
functional derivatives of the Lagrangian functional
as in the definitions (\ref{eq:bigp}), (\ref{eq:smallp}). However,  this
turns out to be quite difficult to do. The problem is that the dependence 
on the velocity $\dot{X}$ of the Lagrangian functional $ L [X, \dot{X},
\ddot{X}]$ is 
difficult to pinpoint. A  more convenient strategy is to consider
the covariant first variation of the action (\ref{eq:haction}) in the 
form (\ref{eq:varp}),
and read out the momenta from the boundary contribution, just as
we did for the DNG model in Sect. 2.

Consider an infinitesimal deformation $X \to X + \delta X$
of the embedding functions for the worldlvolume $m$.
Under this  deformation, to first order, the basic geometric quantities
that characterize the worldvolume $m$ change according to 
\begin{eqnarray}
\delta g_{ab} &=& 2 X_{(a} \cdot \nabla_{b)} \; \delta X\,,
\\
\delta g^{ab} &=& - 2 
X^{(a} \cdot \nabla^{b)} \; \delta X\,,
\\
\delta \sqrt{-g} &=& \sqrt{-g} 
\; g^{ab} \; X_a \cdot \nabla_b 
\; \delta X\,,
\\
\delta K_{ab} &=& - 
n \cdot \nabla_a \nabla_b \; \delta X\,.
\end{eqnarray}
Using  these expressions in the first variation of the 
action (\ref{eq:haction}), we have
\begin{eqnarray}
\fl
\delta S [X] &=& \int_m  [ (\delta \sqrt{-g}) L 
+ \sqrt{-g} ( H_{ab} \delta g^{ab} + L^{ab} \delta K_{ab}) 
\nonumber \\
\fl
&=& \int_m  \; \sqrt{-g}[ L \; g^{ab} \; X_a \cdot 
\nabla_b \delta X - 2 H_{ab} 
X^{a} \cdot \nabla^{b} \delta X
- L^{ab} n \cdot \nabla_a \nabla_b \delta X]\,,
\label{eq:hvar}
\end{eqnarray}
where we have defined
\begin{equation}
H_{ab} = {\partial  L \over \partial g^{ab} }\,, \quad \quad \quad \quad
\quad
L^{ab} = {\partial L \over \partial K_{ab} }\,.
\end{equation}
Note that they are related by the identity \cite{Noether}
\begin{equation}
2 H^{ab} = L^{ac} K_c{}^b + L^{bc} K_c{}^a\,.
\label{eq:id}
\end{equation}
In order to cast
the first variation of the action (\ref{eq:hvar}) in the form 
(\ref{eq:varp}), we integrate (\ref{eq:hvar})
by parts, and we obtain
\begin{equation}
\delta S = \int_m \sqrt{-g} \; 
( E \cdot \delta X 
+  \; \nabla_b Q^b ) \,,
\label{eq:first}
\end{equation}
where we have defined the quantity appearing in the
total divergence
\begin{equation}
Q^b = [ g^{ab} L X_a  
- 2 H^{ab} X_a 
+ \nabla_a (L^{ab} \; n ) ] \cdot \delta X  
- L^{ab} n \cdot \nabla_a \delta X \,. 
\end{equation}
Note that $Q^a$ depends on the infinitesimal deformation $\delta X$ and its 
first derivative. If we had allowed for a dependence of the action on
derivatives of the extrinsic curvature, then $Q^a$ would depend 
also on second derivatives of the deformation.

In (\ref{eq:first}), $E$ is Euler-Lagrange derivative of the action
(\ref{eq:haction})
\begin{eqnarray}
E &=&  - \nabla_b ( g^{ab} L X_a ) + 2 \nabla_b ( H_{ab} X^a )
- \nabla_b \nabla_a ( L^{ab} \, n )
\\
&=& \left( - \nabla_b \nabla_a L^{ab} + K L - L^{ab} K_a{}^c K_{bc}
\right) \, n\,,
\label{eq:eomcov}
\end{eqnarray}
where we have used the Gauss-Weingarten equations for the worldvolume
\begin{eqnarray}
\nabla_a \; X_b &=& - K_{ab} \, n\,, 
\label{eq:gw1} \\
\nabla_a \; n &=& K_{ab} g^{bc} \, X_c\,,
\label{eq:gw2}
\end{eqnarray}
together with the identity (\ref{eq:id}) to obtain the second line.
The Euler-Lagrange derivative is purely normal; this is a consequence
of reparametrization invariance.

For the quadratic model (\ref{eq:haction2}), with
$H_{ab}= K K_{ab}$, and $L^{ab} = K 
g^{ab}$, its Euler-Lagrange derivative takes the form
\begin{equation}
\sqrt{-g} E =   \partial_b \{ 
\sqrt{-g} [ - {1 \over 2}  g^{ab} K^2 X_a  +  2 K  K^{ab}  X_a 
- g^{ab} \nabla_a ( K  n ) ] \}\,, 
\label{eq:eomk2}
\end{equation}
where we have absorbed the factor of $\sqrt{-g}$ in order to facilitate
the comparison with Hamilton's equations later on.

In the canonical analysis, to arrive at the momenta, we  focus on the
boundary term in 
(\ref{eq:first}), involving $Q^a$. We have, using Stokes' 
theorem (\ref{eq:stokes}),  that
\begin{eqnarray}
\fl
\int_m \sqrt{-g} \; \nabla_b Q^b 
&=& -  \int_\Sigma   \sqrt{h} \; \eta_b \; Q^b 
\nonumber \\
\fl
&=& -  \int_\Sigma   
\sqrt{h}  [ L \eta  
-  2  \eta_b H^{ab} X_a  + \eta_b (\nabla_a L^{ab}  n) 
] \cdot \delta X \nonumber \\
&+&  \int_\Sigma \;  \sqrt{h} \; \eta_b L^{ab} n \cdot \nabla_a \delta 
X \,.
\label{eq:bdry}
\end{eqnarray} 
Comparison with 
(\ref{eq:varp}) shows that
the second line will give a contribution to the canonical momenta $p$
conjugate to $X$. 
However,
the dependence on $\delta \dot{X}$ is only implicit  in the last term
of (\ref{eq:bdry}). In order to isolate it, we
consider an arbitrary worldvolume
vector $V^a$,  and we  exploit the completeness relation on the worldvolume
(\ref{eq:compli})  
to get
\begin{eqnarray}
\fl
V^a n \cdot \nabla_a \delta X 
&=& V_a g^{ab} n \cdot \nabla_b \delta X 
\nonumber \\
\fl
&=& - V_a \eta^a \eta^b n \cdot \nabla_b \delta X
+ V_b X^{bA} n \cdot \partial_A \; \delta X
\nonumber \\
\fl
&=& 
- {1 \over N} V_b \eta^b n \cdot \delta \dot{X}
+ \left( {1 \over N} V_b \eta^b N^A + V_b X^{bA}\right) n \cdot 
\partial_A \; \delta X\,. \end{eqnarray}
We now set $V^a = \sqrt{h} \eta_b L^{ab}$, and use this expression
in the last term of (\ref{eq:bdry}) to obtain
\begin{eqnarray}
\fl
\int_\Sigma   &\sqrt{h}& \eta_b L^{ab} n \cdot \nabla_a \delta X 
\nonumber \\
\fl
&=& \int_\Sigma \left[ 
- {\sqrt{h} \over N} \eta_b  L^{ab} \eta_a n \cdot  \delta \; \dot{X}
+ {\sqrt{h} \over N}  \left( \eta_a L^{ab} \eta_b  N^A + N \eta^a L_{ab} 
X^{bA}\right) n \cdot \partial_A \delta X \right]
\nonumber \\
\fl
&=& \int_\Sigma \left\{ 
- {\sqrt{h} \over N} \eta_b  L^{ab} \eta_a n \cdot  \delta \dot{X}
- \partial_A \left[ {\sqrt{h} \over N}  \left( \eta_a L^{ab} \eta_b  N^A + N 
\eta^a L_{ab} X^{bA}\right)  n  \right]\cdot  \delta X \right\}.
\label{eq:601}
\end{eqnarray}
We have integrated by parts and disregarded a total spatial divergence.
We see that the boundary term which involves a derivative of the
variation identifies the momenta $P$ conjugate to $\dot{X}$ 
together with
a contribution to the momenta $p$ given by a spatial divergence.
The appearance of this latter term is a novelty introduced by the
Hamiltonian formulation.

Using (\ref{eq:601}) in (\ref{eq:bdry}) , comparison with (\ref{eq:varp})
produces the canonical momenta $P, p$ conjugate to $\dot{X}, X$,
respectively, as
\begin{equation}
P = - {\sqrt{h} \over N} \; \eta_b  L^{ab} \eta_a \; n\,,
\label{eq:59}
\end{equation}
\begin{equation}
\fl
p = - \sqrt{h} [
L \eta  
-  2  \eta_b  H^{ab} X_a 
+ \eta_b \nabla_a ( L^{ab} \, n ) ] 
+ \partial_A [ N^A P - \sqrt{h} \eta^a L_{ab} 
X^{bA} \,  n  ]\,.
\label{eq:60}
\end{equation}
The momenta $P$ conjugate to $\dot{X}$ is always normal to the worldvolume,
and it vanishes if the action does not depend on the extrinsic curvature.
The momenta $p$ conjugate to $X$ is given by two parts, with 
the second a spatial divergence. The first part is related to the
conserved linear momentum for this system. We will 
analyze this relationship  in the next section.
Note that both $P$ and $p$ are spatial densities of weight one;
their spatial derivative is independent of the spatial affine
connection. This fact will be used extensively below.
Note also how the case of a DNG model is quite special: only the
first term in (\ref{eq:60}) survives, and $p$ is purely tangential
to the worldvolume.

The momenta $p$ conjugate to $X$ can be written in an alternative way
using the Gauss-Weingarten equation (\ref{eq:gw2})
together with the identity (\ref{eq:id}), as
\begin{equation}
\fl
p = - \sqrt{h} [
L \eta  
-   \eta_b  L^{ac} K^{b}{}_c X_a 
+ \eta_b (\nabla_a  L^{ab} ) \, n  ] 
+ \partial_A [ N^A P - \sqrt{h} \eta^a L_{ab} 
X^{bA} \,  n  ]\,.
\end{equation}

If we specialize our considerations to the quadratic model  
(\ref{eq:haction2}), 
then
the momenta (\ref{eq:59}), (\ref{eq:60}) take the form
\begin{eqnarray}
P &=&  {\sqrt{h} \over N} \; K \;  n\,,
\label{eq:P21}
\\
p &=&  \sqrt{h} \left[ - {1 \over 2} K^2 \eta  +  2 \eta_b K K^{ab} X_a 
- \eta^a \nabla_a  ( K  n )\right] 
+ \partial_A \left( N^A P \right)\,.
\label{eq:P22}
\end{eqnarray}
This concrete example makes it clear how the higher derivative nature
of the model determine the structure of the canonical momenta $P,p$.

\section{Conservation laws}

The conservation laws associated with the
Poincar\'e invariance for geometric models of relativistic
objects were examined in \cite{Noether}, exploiting Noether's theorem.
This made it possible to derive general expressions for both the
linear and angular momenta for any action of the form (\ref{eq:haction})
depending on
the extrinsic curvature. 
In this section , we examine the relationships between
these conserved quantities and the canonical momenta.

Return to the first variation of the action in the form 
(\ref{eq:first}). If we specialize the variation of the embedding functions
to an infinitesimal constant translation $\delta X  =
\epsilon$ and we set $Q^a = - \epsilon_\mu T^{\mu\,a}$
(we refer to \cite{Noether} for details), we obtain for the
energy-momentum tensor associated with the action (\ref{eq:haction}), 
\begin{equation}
T^{\mu\,a}  = (L g^{ab} -  L^{bc} K^{a}{}_c ) X^\mu_b
+ \nabla_b ( L^{ab}) n^\mu \,.
\end{equation}
In particular, for an action that is invariant under
translations, it follows that we can write the
Euler-Lagrange derivative as
\begin{equation}
E^\mu = \nabla_a T^{\mu\, a}\,.
\end{equation}

Let us define the linear momentum density associated with the object 
 $\Sigma$ by
\begin{equation}
\pi^\mu = \sqrt{h} \eta_a T^{\mu\, a}\,.
\end{equation}
Then we see that it differs from the canonical momenta $p$ 
conjugate to $X$ (\ref{eq:60}) 
by a total spatial divergence,
\begin{equation}
p = \pi + 
 \partial_A [ N^A P - \sqrt{h} \eta^a L_{ab} 
X^{bA} \,  n  ]\,.
\end{equation}
In particular, it follows that both $p$ and $\pi$ give the same
total linear momentum of the object $\Sigma$ since
\begin{equation}
\int_\Sigma p = \int_\Sigma \pi\,,
\end{equation}
up to a vanishing boundary term.

The boundary term is related to the angular momentum 
asociated with the action (\ref{eq:haction}). 
If we specialize the variation of the embedding functions
to an infinitesimal Lorentz transformation $\delta X^\mu  =
\omega^\mu{}_\nu X^\nu$ , with $\omega_{\mu\nu} = - \omega_{\nu\mu}$, 
and we set $Q^a = - \omega_{\mu \nu} M^{\mu\nu\, a}$, we obtain for the
angular momentum
\begin{eqnarray}
M^{\mu\nu\, a} 
&=& 
{1 \over 2} [ T^{\mu\,a} X^\nu + L^{ab} n^\mu X^\nu_b
- (\mu \leftrightarrow \nu ) ]\nonumber \\
&=& 
{1 \over 2} [ T^{\mu\,a} X^\nu 
- L^{ab} \eta_b n^\mu \eta^\nu
+ L^{ab} X_b^A n^\mu X^\nu_A
- (\mu \leftrightarrow \nu ) ]\,,
\end{eqnarray}
where in the second line we have decomposed $X_b$ along the basis
$\{ \eta , X_A \}$. The angular momentum is given by the
sum of an orbital part, that depends on the origin, and
a differential part. The latter is related to the boundary
term that appears in (\ref{eq:60}).

\section{Hamiltonian and constraints}

In this section, we restrict our attention to the  model quadratic 
in the mean extrinsic curvature
(\ref{eq:haction2}), since the generic case given by (\ref{eq:haction})
can be treated along the same lines.  We derive the canonical
Hamiltonian (\ref{eq:hc1}) and we identify the constraints
on the phase space variables that follow from the reparametrization
invariance of the action.

The canonical Hamiltonian is given by the Legendre transformation
with respect to both $\ddot{X}$ and $\dot{X}$ (see (\ref{eq:hc1})),
\[
H_c [X,p;\dot{X}, P ] 
= \int_\Sigma  ( P \cdot \ddot{X} + p \cdot \dot{X} ) - 
L [ X, \dot{X} , \ddot{X} ]\,.
\]
Using (\ref{eq:P21}) for the higher momenta $P$, we have
immediately that the Lagrangian functional can be expressed in terms
of the phase space variables as 
\begin{equation}
L [ X, \dot{X} , \ddot{X} ] = {1 \over 2} \int_\Sigma N \sqrt{h} K^2 
= {1 \over 2} \int_\Sigma {N^3 \over \sqrt{h}} P^2 \,. 
\label{eq:69}
\end{equation}
In order to carry out the Legendre transformation, we need to express
the acceleration $\ddot{X}$ in terms of $\{ P , \dot{X}, X \}$. Recall that 
$p$ is left alone.  
With (\ref{eq:P21}) for the momenta $P$, we have
\begin{equation}
P \cdot \ddot{X}
= {\sqrt{h} \over N} K n \cdot \ddot{X}\,.
\label{eq:77}
\end{equation}
We need to write the normal component of the acceleration in
terms of the extrinsic curvature.
For this purpose, note that it is given by (minus) the
time-time projection of the extrinsic curvature
\begin{equation}
K_{00} = \dot{X}^a \dot{X}^b K_{ab} =
- n \cdot \ddot{X}\,.
\label{eq:pr1}
\end{equation}
The other projections of the extrinsic curvature 
are
\begin{eqnarray}
K_{0A} &=& \dot{X}^a X^b_A K_{ab} = - n \cdot \partial_A \dot{X}\,,
\label{eq:pr2}
\\
K_{AB} &=& X^a_A X^b_B K_{ab} = - n \cdot  {\cal D}_A {\cal D}_B X\,.
\label{eq:pr3}
\end{eqnarray}

We  use the definition of the velocity $\dot{X}$ (\ref{eq:velo}), together
with the worldvolume completeness relation (\ref{eq:compli}) to obtain
for the time-time projection of the extrinsic curvature (\ref{eq:pr1}),
\begin{eqnarray}
K_{00} &=& 
N^2 \eta^a \eta^b K_{ab}
+ 2 N N^A \eta^a X^b_A K_{ab}
+ N^A N^B  K_{AB} \nonumber \\
&=& - 
N^2 K + 2 N N^A \eta^a X^b_A K_{ab}
+ ( N^2 h^{AB} + N^A N^B) K_{AB} \nonumber \\
&=& - N^2 K + 2 N N^A K_{0A}
+ ( N^2 h^{AB} - N^A N^B) K_{AB}\,,
\label{eq:k00}
\end{eqnarray}
therefore from (\ref{eq:77}) we have 
\begin{equation}
P \cdot \ddot{X} = 
{\sqrt{h} \over N} [ 
N^2 K^2  - 2  N^A   K K_{0A}
-  ( N^2 h^{AB} - N^A N^B) K K_{AB}]\,,
\end{equation}
and using once more (\ref{eq:P21}) for $P$, we find that $P \cdot
\ddot{X}$ is 
expressed in terms of the phase space variables by
\begin{equation}
P \cdot \ddot{X} = {N^3 \over \sqrt{h}} P^2 
+ 2 N^A P \cdot \partial_A \dot{X} 
+ (N^2 h^{AB} - N^A N^B ) P \cdot {\cal D}_A {\cal D}_B X\,.
\end{equation}
Using (\ref{eq:69}) for the Lagrangian functional and this last 
expression in the Legendre transformation (\ref{eq:hc1}) gives the
Hamiltonian as 
\begin{equation}
\fl
H_c [X,p;\dot{X}, P ] 
= 
\int_\Sigma \left[ p \cdot \dot{X} +
{N^3 \over 2 \sqrt{h}} P^2 
+ 2 N^A P \cdot \partial_A \dot{X} 
+ (N^2 h^{AB} - N^A N^B ) P \cdot {\cal D}_A {\cal D}_B X \right]\,.
\label{eq:hc2}
\end{equation}
The dependence of this Hamiltonian on $P, p, \dot{X}$ is explicit.
On the other hand, the dependence on $X$ is not so evident.
Besides the dependence on $X$ through the induced metric on $\Sigma$,
$h_{AB}$, and the last term, where $X$ appears explicitly, there
is also a dependence on $X$ of both the lapse function and the
shift vector $N , N^A$. We will address this technical point explicitly in
the next section.

Note that for a point-like object, this Hamiltonian specializes to  
\begin{equation}
H_c [X,p;\dot{X}, P ] 
= p \cdot \dot{X} +
{N^3 \over 2 } P^2 \,.
\end{equation}
This agrees with the Hamiltonian obtained {\it e.g.} in \cite{HFS}
for  a relativistic particle model with an action quadratic in the geodesic
curvature.

Just as in the case of a DNG extended object, reparametrization 
invariance  of the action (\ref{eq:haction2}) implies the existence of 
constraints. 
According to the Dirac-Bergmann theory of higher derivative theories,
we immediately identify the primary
constraints that involve the highest momenta $P$ from its definition
as
\begin{eqnarray}
C &=& P \cdot \dot{X} = 0 \,,
\label{eq:c1}
\\
C_A &=& P \cdot X_A = 0\,.
\label{eq:c2}
\end{eqnarray}
There are no other primary constraints involving $P$,
as one can check by computing the Hessian (\ref{eq:hess})
\begin{equation}
H_{\mu \nu} = {1 \over N^4} n_\mu n_\nu\,,
\end{equation}
The only null eigenvectors are $\dot{X}$ and $X_A$.
The quadratic model (\ref{eq:haction2}) is representative 
of the generic model (\ref{eq:haction}) as long as 
this is the case.

The Hamiltonian that generates the dynamics is given by adding
the primary constraints (\ref{eq:c1}), (\ref{eq:c2}) to the
canonical Hamiltonian (\ref{eq:hc2}),
\begin{equation}
H = H_c + \int_\Sigma ( \lambda C + \lambda^A C_A )\,,
\label{eq:hext}
\end{equation}
with $\lambda, \lambda^A$  arbitrary  Lagrange multipliers.

There are also secondary constraints. They do not participate
in the determination of the dynamics, but they are important
as generators of worldvolume reparametrizations.
From the conservation in time of the primary constraint
(\ref{eq:c1}), we obtain the secondary constraint that
the integrand of the canonical Hamiltonian (\ref{eq:hc2}) must vanish,
\begin{equation}
S = {\cal H}_c = 0\,.
\label{eq:s1}
\end{equation}
where we have defined $H_c = \int_\Sigma {\cal H}_c$.
Just as in the DNG case, this was to be expected on the
basis of reparametrization invariance. 
Conservation in time of the other primary constraint (\ref{eq:c2})
produces the additional secondary constraint
\begin{equation}
S_A = p \cdot X_A + P \cdot \partial_A \dot{X} =0\,.
\label{eq:s2}
\end{equation}
There are no tertiary constraints. The constraint algebra of
the four constraints is in involution; they are first class. 
The full Poisson algebra of the constraints is given in Appendix B.

The constraint (\ref{eq:s2}) is the generator of
diffeomorphisms tangential to $\Sigma$, as it can be
verified directly considering its Poisson bracket with the
phase space variables (see Appendix B).
The vanishing of the canonical Hamiltonian, the secondary constraint
(\ref{eq:s1}), generates diffeomorphisms out of
$\Sigma$ onto the worldvolume $m$. Again, this can be verified by
considering the Poisson bracket of $S$ with the phase space variables.
This can be read out from the Hamilton's equations of motion given below
in Sect. 7. Also the effect of the primary constraints on the
phase space variables is listed in Appendix B.

The number of physical degrees of
freedom for the action (\ref{eq:haction2}) is equal to 
$1/2[4 (d+2) - 4(d+1)]= 2$. 
There are two physical degrees
of freedom associated with the single normal to
the worldvolume. For higher co-dimension there will be two physical
degree of freedom for each normal. This is twice the number of physical
degrees of freedom in the DNG model. The appearance of extra degrees of
freedom is a generic feature of  higher derivative theories.

\section{Hamilton's equations}

In this section we consider the Hamilton equations for the 
model (\ref{eq:haction}), generated by the Hamiltonian (\ref{eq:hext}). 
We check explicitly that, at the end of the day, Hamilton's equations
correctly reproduce the equations of motion.

As we pointed out in Sect. 3, the first Hamilton equation 
$\dot{X} = \delta H / \delta p$ is an 
identity, since the only dependence on $p$ is thought the
$p \cdot \dot{X}$ term in $H_c$. The second Hamilton equation
reproduces
the form of the momenta $P$ given by (\ref{eq:P21}), and
furthermore identifies the 
form of
the Lagrange multipliers $\lambda, \lambda^A$ in terms of the canonical
variables. We obtain
\begin{equation}
\fl
\ddot{X} =
{\delta H \over \delta P}
= {N^3 \over \sqrt{h}} P 
+ 2 N^A \partial_A \dot{X} 
+ (N^2 h^{AB} - N^A N^B ) {\cal D}_A {\cal D}_B X 
+ \lambda \dot{X} + \lambda^A X_A\,.
\label{eq:H2}
\end{equation}
To see that this gives (\ref{eq:P21}), the quickest way is to
contract with the worldvolume normal $n$, obtaining
\begin{eqnarray}
P \cdot n &=& {\sqrt{h} \over N^3 } [ - K_{00}
+ 2 N^A K_{0A}
+ (N^2 h^{AB} - N^A N^B ) K_{AB}  \nonumber \\
&=& {\sqrt{h} \over N} K\,,
\end{eqnarray}
where we have used (\ref{eq:k00}).
Contraction of (\ref{eq:H2}) with $\dot{X}$ and $X_A$, and use of the 
constraints identifies the Lagrange multipliers as (see Appendix A)
\begin{eqnarray}
\lambda &=& {\cal D}_A N^A - {N^2 \over \sqrt{h}} 
\eta^a \nabla_a \left( {\sqrt{h} \over N}\right)\,,
\label{eq:82}
\\
\lambda^A &=&   {N^2 \over \sqrt{h}}
\eta^a \nabla_a \left( {N^A \sqrt{h} \over N} \right)
- N^A {\cal D}_B N^B + N h^{AB} {\cal D}_B N\,.
\label{eq:83}
\end{eqnarray}
Note that these relationships involve derivatives of the lapse function, the
shift vector and the $\Sigma$ volume element. They also involve
mixing: $\lambda$ and $\lambda^A$ depend on both $N$ and $N^A$.
Gauge fixing will amount to a specification of 
$\lambda$ and $\lambda_A$ or,
equivalently, $N$ and $N^A$.
Clearly, this  will be qualitatively very different 
from the familar DNG case.

Let us turn to the third Hamilton equation. This identifies the form 
of the momenta $p$ as given by (\ref{eq:P22}). We have
\begin{eqnarray}
\fl
\dot{P} = &-& {\delta H  \over \delta \dot{X}} = 
- p + {3 \over 2} {N^2 \over \sqrt{h}} P^2 \eta
- 2 h^{AB} ( P \cdot \partial_A \dot{X} ) X_B + 2 \partial_A ( N^A P ) 
\nonumber \\
\fl
&+& 2 N h^{AB} (P \cdot {\cal D}_A {\cal D}_B X ) \eta
+ 2 N^A h^{BC} (P \cdot {\cal D}_A {\cal D}_B X ) X_C
- \lambda P\,. 
\label{eq:hthird}
\end{eqnarray}
To see that this expression coincides with  (\ref{eq:P22}), first note that
\begin{eqnarray}
\fl
{3 \over 2} {N^2 \over \sqrt{h}} P^2 \eta
&-& 2 h^{AB} ( P \cdot \partial_A \dot{X} ) X_B 
+ 2 N h^{AB} (P \cdot {\cal D}_A {\cal D}_B X ) \eta
+ 2 N^A h^{BC} (P \cdot {\cal D}_A {\cal D}_B X ) X_C
\nonumber \\
\fl
&=& \sqrt{h} \left[ ({3 \over 2} K^2 - 2 K h^{AB} K_{AB} ) \eta
+ {2 \over N} ( h^{AB} K K_{0A} - N^A h^{BC} K K_{AC} ) X_B \right]
\nonumber \\
\fl
&=& - {1 \over 2} \sqrt{h} K^2 \eta + 2 \sqrt{h}
K K^{ab} \eta_a X_b\,,
\label{eq:83a}
\end{eqnarray}
where we have used (\ref{eq:g0b}), (\ref{eq:gAb}).
Therefore we find
\begin{equation}
p =  - {1 \over 2} \sqrt{h} K^2 \eta + 2 \sqrt{h}
K K_{ab} \eta_a X_b 
- \dot{P} + 2 \partial_A ( N^A P ) - \lambda P\,,
\end{equation}
and inserting the form of the Lagrange mutiplier (\ref{eq:83})
we recover (\ref{eq:P22}) for $p$.

The fourth Hamilton equation is
\begin{equation}
\dot{p} = - {\delta H \over \delta X}\,.
\label{eq:87}
\end{equation}
Using the explicit form of the momenta and of the Lagrange 
multipliers, this equation ought to reproduce the equation of motion
(\ref{eq:eomk2}). To show this turns out to be a little more involved
than one may have expected. First of all,
in the variation of the Hamiltonian with respect to $X$,
it is necessary to recognize that, besides the obvious dependence
on $X$ of the metric induced on $\Sigma$, $h_{AB}$,  there
is also a dependence implicit in the lapse function $N$ and
the shift vector $N^A$.
Holding $\dot{X}$ fixed, we have that the unit normal to $\Sigma$ onto
$m$ varies with respect to $X$ as
\begin{equation}
\delta \eta = - 
( \eta \cdot \partial_A \delta X ) h^{AB} X_B - {N^A \over N}
( n \cdot \partial_A \delta X ) n\,.
\end{equation}
It implies that
the variation of the lapse function with respect to $X$, 
again holding $\dot{X}$ fixed, is
\begin{equation}
\delta N = 
N^A ( \eta \cdot \partial_A \delta X )\,.
\end{equation}
For the variation of the shift vector with respect to the embedding 
functions $X$ for $\Sigma$, we find 
\begin{equation}
\delta N^A = N h^{AB} ( \eta \cdot \partial_B \delta X )
- N^B h^{AC} ( X_C \cdot \partial_B \delta X )\,.
\end{equation}

Consider now the variation of the Hamiltonian, holding $\dot{X}, p , P$ 
fixed. We have
\begin{eqnarray}
\fl
\delta H &=& \int_\Sigma \left\{ \left[ - 
{N^3 \over 2 \sqrt{h}} P^2
h^{AB} - 2 N^A h^{BC} 
(P \cdot \partial_C \dot{X} ) 
- 2 N^2 h^{AC} h^{BD} 
( P \cdot {\cal D}_C {\cal D}_D X ) \right. \right. \nonumber \\
\fl
&+& \left. 2 N^A N^D h^{BC} 
( P \cdot {\cal D}_C {\cal D}_D X ) \right]
( X_B \cdot \partial_A \delta X )
\nonumber \\
\fl
&+& \left[ {3 N^3 \over 2 \sqrt{h}} N^A P^2
+ 2 N h^{AB} (P \cdot \partial_B \dot{X} ) 
+ 4 N N^{(A} h^{B)C}  
( P \cdot {\cal D}_B {\cal D}_C X ) \right. \nonumber \\
\fl
&-& \left. 4 N N^{(B}  h^{A)C}( P \cdot {\cal D}_C {\cal D}_B X ) \right] 
( \eta \cdot \partial_A \delta X )
\nonumber \\
\fl
&+& \left. (N^2 h^{AB} - N^A N^B ) ( P \cdot {\cal D}_A {\cal D}_B  \delta 
X ) + \lambda^A P \cdot \partial_A \delta X \right\}\,.
\label{eq:91}
\end{eqnarray}
We integrate by parts and, disregarding a total spatial divergence, we 
carry out the functional derivative in (\ref{eq:87}), so that we obtain
\begin{eqnarray}
\fl
\dot{p} &=& - \partial_A \left\{ \left[ 
{N^3 \over 2 \sqrt{h}} P^2
h^{AB} + 2 N^A h^{BC} 
(P \cdot \partial_C \dot{X} ) 
+ 2 N^2 h^{AC} h^{BD} 
( P \cdot {\cal D}_C {\cal D}_D X ) \right. \right. \nonumber \\
\fl
&-& \left. 2 N^A N^D h^{BC} 
( P \cdot {\cal D}_C {\cal D}_D X ) \right] X_B 
\nonumber \\
\fl
&+& \left[ - {3 N^3 \over 2 \sqrt{h}} N^A P^2
- 2 N h^{AB} (P \cdot \partial_B \dot{X} ) 
- 4 N N^{(A} h^{B)C}  
( P \cdot {\cal D}_B {\cal D}_C X ) \right. \nonumber \\
\fl
&+& \left. 4 N N^{(B}  h^{A)C}( P \cdot {\cal D}_C {\cal D}_B X ) \right]
\eta  
\nonumber \\
\fl
&+& \left. 
\partial_B [ (N^2 h^{AB} - N^A N^B ) P ] -
\lambda^A P \right\}\,.  
\end{eqnarray}
Now, we need to recognize that the part proportional to the
shift vector $N^A $ is related to the momenta $p$ via (\ref{eq:83}),
so that the equations of motion can be written in canonical form as
\begin{eqnarray}
\fl
\dot{p} &=& \partial_A \left\{ N^A p + N^A \dot{P} - 2 N^A \partial_B
( N^B P )
+ \lambda N^A P + \lambda^A P
+ {\cal D}_B [ (N^2 h^{AB} - N^A N^B ) P ] \right. \nonumber \\
\fl
&+& N^2 \left[ {N \over 2 \sqrt{h}}
P^2 h^{AB} + 2  h^{AC} h^{BD} 
(P \cdot {\cal D}_C {\cal D}_D X )  
\right] X_B \nonumber \\
\fl 
&+& \left. 2 N \left[ N^C h^{AB} ( P \cdot {\cal D}_B {\cal D}_C X )
- ( P \cdot \partial_B \dot{X} ) h^{AB} \right] \eta \right\}\,.
\label{eq:93}
\end{eqnarray}

It is not at all obvious that this form of the equations of motion 
coincides with the covariant expression  
(\ref{eq:eomk2}). In order to be able to compare the two, 
we decompose the (densitized) Euler-Lagrange derivative in (\ref{eq:eomk2}) 
with respect to the worldvolume
basis $\{ \dot{X}, X_A \}$.
This gives
\begin{eqnarray} 
\fl
\sqrt{-g} E &=&   \partial_0 \left\{ 
N \sqrt{h} \left[ - {1 \over 2}  g^{0a} K^2 X_a  +  2 g^{0c} K  K^{a}{}_c  
X_a - g^{0a} \nabla_a ( K  n ) \right] \right\}
\nonumber \\
\fl
&+& \partial_A \left\{ 
N \sqrt{h} \left[ - {1 \over 2}  g^{Aa} K^2 X_a  +  2  g^{Ac} K  K^{a}{}_c  
X_a - g^{Aa} \nabla_a ( K  n ) \right] \right\}\,,
\end{eqnarray}
and we  use (\ref{eq:g0b}), (\ref{eq:gAb}) 
for $g^{0a}, g^{Ab}$, to obtain
\begin{eqnarray}
\fl
\sqrt{-g} E
&=& \partial_0 \left\{ \sqrt{h} \left[  {1 \over 2} K^2 \eta  -  2 \eta^c K  
K^{a}{}_c  X_a + \eta^a \nabla_a ( K  n ) \right] \right\}
\nonumber \\
\fl
&+& \partial_A \left\{ 
\sqrt{h} \left[ - {N \over 2}  K^2 h^{AB} X_A
- {1 \over 2} N^A K^2 \eta
  +  2  N h^{AB}  K  K^{a}{}_B  X_a 
+ 2 N^A \eta^c K K_a{}^c X_c \right. \right. \nonumber \\
\fl
&-& \left. \left. N h^{AB} {\cal D}_B ( K n )
- N^A \eta^a \nabla_a ( K n ) \right] \right\} \,.
\end{eqnarray}
The argument of the time derivative is
related to $p$, using (\ref{eq:P22}), and the same combination
appears in the spatial divergence, in the terms proportional to $N^A$,
so that 
\begin{eqnarray}
\fl
\sqrt{-g} E &=& - \dot{p} + \partial_0 \partial_A ( N^A P )
+ \partial_A \left[ N^A p  -  N^A \partial_B (N^B P ) -
{N \sqrt{h} \over 2} 
K^2 h^{AB} X_B \right. \nonumber \\
\fl
&-&  
N \sqrt{h} h^{AB} {\cal D}_B ( K n )
- 2 N \sqrt{h} K K^{Aa} \eta_a \eta 
+ 2 N \sqrt{h} K K^{AB} X_B 
\nonumber \\
\fl 
&-& \left. 
N \sqrt{h} {\cal D}^A (K n)\right]\,,
\end{eqnarray}
which in terms of the phase space variables takes the form
\begin{eqnarray}
\fl
\sqrt{-g} E = &-& \dot{p} + \partial_A \{ 
\partial_0 ( N^A P ) + N^A p - N^A \partial_B ( N^B P ) - N \sqrt{h}
h^{AB} {\cal D}_B ( K n ) 
\nonumber \\
\fl
&-& N^2 \left[ {1 \over 2} {N \over \sqrt{h}} P^2  h^{AB}
+ 2 h^{AC} h^{BD} ( P \cdot {\cal D}_C {\cal D}_D X ) \right] X_B
\nonumber \\
\fl
&+& 2 N \left[( P \cdot \partial_B \dot{X} ) h^{AB}   -  N^B ( p \cdot {\cal 
D}_A {\cal D}_B X ) \right] \eta \,.
\end{eqnarray}
To see that the vanishing of this expression coincides with
the canonical equations of motion (\ref{eq:93}) it is sufficient to use the
explicit form of the Lagrange multipliers (\ref{eq:82}), 
(\ref{eq:83}) in 
(\ref{eq:93}).

\section{Higher co-dimension}

In order to minimize the already vexing amount of formalism, so far we
have 
restricted our attention to the case
where the worldvolume $m$ is a timelike  hypersurface in spacetime,
with a single normal vector field. In this
section, we lift this restriction, and consider arbitrary co-dimension. 
As it becomes clear soon enough, 
the modification in the Hamiltonian formulation consists basically of a
decoration with
normal indices every time a normal vector field appears.

In a fixed Minkowski background of dimension $d+N+1$, for a relativistic
extended object of dimension $d$, we have $N$ unit spacelike normal
vector fields $n^i$ $(i,j,\dots = 1,2, \cdots, N)$, defined implicitly by
$ n^i \cdot X_a = 0$ and $ n^i \cdot n^j = \delta^{ij}$, with $\delta^{ij}$
the Kronecker delta.
The Gauss-Weingarten equations take the form
\begin{eqnarray}
\nabla_a X_b &=& - K_{ab}{}^i n_i\,,
\\
\nabla_a n^i &=& K_{ac} g^{bc} X_c + \omega_a{}^{ij} n_j\,.
\end{eqnarray}
There are $N$ extrinsic curvature tensors, one along each normal, and
$N$ mean extrinsic curvatures, $K^i = g^{ab} K_{ab}{}^i$.
The  new structure is the normal connection $\omega_a{}^{ij} = -
\omega_a{}^{ji}$ associated with the $O(N)$ invariance under
rotations of the normal vector fields. We use it to define the
$O(N)$ covariant derivative $\tilde\nabla_a = \delta^{ij} \nabla_a
- \omega_a{}^{ij}$. As long as one is concerned with quantities
invariant under rotations of the normals, the normal connection
will appear only in intermediate steps of the calculations.

We consider an action that depends on the $N$ extrinsic curvatures,
generalizing (\ref{eq:haction}) to
\begin{equation}
S [X] = \int_m \sqrt{-g} L ( g^{ab}, K_{ab}{}^i )\,,
\label{eq:hcd}
\end{equation}
which, besides the requirements of Poincar\'e and reparametrization 
invariance, is also subject to the requirement of invariance  under 
rotations of the 
normal fields, such as {\it e.g.} the higher co-dimension version of the
quadratic model (\ref{eq:haction2}), 
\begin{equation}
S [X] = {1 \over 2} \int_m \sqrt{-g} K^i K_i\,.
\label{eq:hcd2}
\end{equation}
For the case of a relativistic string, this action was proposed by
Polyakov, and
independently by Kleinert \cite{Polyakov,Kleinert}.

Let us describe briefly how the Hamiltonian formulation
generalizes to higher co-dimension. 
For an action of the form (\ref{eq:hcd}), the momenta are
\begin{equation}
P = - {\sqrt{h} \over N} \; \eta_b  L^{ab}{}_i \eta_a \; n^i\,,
\end{equation}
\begin{equation}
\fl
p = - \sqrt{h} [
L \eta  
-  2  \eta_b  H^{ab} X_a 
+ \eta_b \nabla_a ( L^{ab}_i \, n^i ) ] 
+ \partial_A [ N^A P - \sqrt{h} \eta^a L_{ab\, i} 
X^{bA} \,  n^i  ]\,.
\end{equation}
where $L^{ab}_i = \partial L / \partial K_{ab}{}^i $.
For the quadratic model (\ref{eq:hcd2}), the momenta become
\begin{equation}
P = {\sqrt{h} \over N} \; K_i \; n^i\,,
\end{equation}
\begin{equation}
p = - \sqrt{h} \left[
{1 \over 2} K^i K_i \eta  
-  2  \eta_b  K^i K^{ab}{}_i X_a 
+ \eta^a \nabla_a ( K_i \, n^i ) \right] 
+ \partial_A ( N^A P )\,.
\end{equation}
The canonical Hamiltonian is unchanged with respect to the
co-dimension one case analyzed in Sect. 6, and it is given
by (\ref{eq:hc2}), also the phase space constraints, their
algebra and the Hamilton's equations are unchanged.
What will change is the counting of the physical degrees
of freedom of the theory; in the non-degenerate case there
will be two physical dgrees of freedom for each normal.

\section{Concluding remarks}

In this paper, we have presented a Hamiltonian formulation of the dynamics
of a relativistic extended object propagating in Minkowski
spacetime described by a local action that
depends on the extrinsic curvature of its worldvolume.
We have focused our attention on a model quadratic in the mean extrinsic
curvature, in the special case of an object of dimension $d$
propagating in a Minkowski background of dimension $d+2$. However, 
this is not a real restriction: an arbitrary 
dependence on curvature and a higher co-dimension requires
only minimal and straightforward modifications.
(The generalization to an 
arbitrary background  can be carried out along similar lines.)
This is true as long as the structure of the primary constraints
is the same as in the quadratic case. One obvious case where
this does not happen is for models with a linear dependence on the
acceleration, such as the Einstein-Hilbert action. Such special
cases will be considered elsewhere \cite{CGREH}. A less obvious 
special case is provided by the `gonihedric' string model, with
action \cite{Savvidy}
\begin{equation}
S [X] = \int_m \sqrt{-g} \sqrt{ K^i K_i }\,.
\end{equation} 
For this model, there are additional primary constraints and the
study of its Hamiltonian formulation is in progress \cite{Savham}.

\ack
JG thanks Denjoe O' Connor
for hospitality during his stay at DIAS. He also acknowledges 
partial support from DGAPA-PAPIIT grant IN114302.
ER acknowledges partial support from CONACyT under
grant CO1-41639 and PROMEP-2003.

\vspace{1cm}

\noindent{\bf APPENDIX A: USEFUL FORMULAE}

\vspace{.5cm}

In this Appendix we collect some useful formulae about the geometries
of the worldvolume $m$, its spatial slice $\Sigma$, and their
relationships.
The Gauss-Weingarten equations for the worldvolume $m$ are
\begin{eqnarray*}
\nabla_a X_b &=& -K_{ab}  n \,, \\
\nabla_a n &=& K_{ab} g^{bc}\,.
\end{eqnarray*}
Their integrability conditions give 
the Gauss-Codazzi-Mainardi equations for the worldvolume, in a Minkowski
background, 
\begin{eqnarray}
{\cal R}_{abcd} &=& K_{ac} K_{bd} - K_{ad} K_{bc}\,, \\
\nabla_a K_{bc} &=& \nabla_b K_{ac}\,.
\end{eqnarray}
Contractions with the inverse induced metric $g^{ab}$ gives
\begin{eqnarray}
{\cal R}_{ab} &=& K K_{ab} - K_{a}{}^c K_{bc}\,, \\
{\cal R} &=& K^2  - K^{ab} K_{ab}\,, \\
\nabla_b K^b{}_{a} &=& \nabla_a K\,.
\end{eqnarray}
The Gauss-Weingarten equations for $\Sigma$ seen as embedded
in the worldvolume $m$ are
\begin{eqnarray}
{\cal D}_A X_B &=&  k_{AB} \eta \,, \\
{\cal D}_A \eta &=& k_{AB} h^{BC} X_C\,,
\end{eqnarray}
where the spatial tensor
$k_{AB}$ is the extrinsic curvature of $\Sigma$ as embedded in the
worldvolume $m$,
\begin{equation}
k_{AB} = - \eta \cdot {\cal D}_A {\cal D}_B X\,.
\end{equation} 
The projection of the Gauss-Weingarten equations along the
basis $\{ \eta , n, X_A \}$ adapted to $\Sigma$ gives
\begin{eqnarray}
\fl
\ddot{X} &=& - K_{00} \,n +
\left( \dot{N}_A + N {\cal D}_A N -
N^B {\cal D}_A N_B \right)\, h^{AC} X_C 
\nonumber \\
\fl
&+& \left( \dot{N} + N^A {\cal D}_A N +
N^A N^B k_{AB} \right)\,\eta \,,
\label{eq:acceleration1}
\\
\fl
\partial_A \dot{X} &=& ({\cal D}_A N  + N^Bk_{AB})\,\, \eta +
({\cal D}_A N_B + N k_{AB}) h^{BC} X_C
- K_{0A} n \,, \\
\fl
{\cal D}_A {\cal D}_B X &=& k_{AB}\,\eta - K_{AB} n \,,   \\
\fl
\dot{\eta } &=& ({\cal D}_A N + N^B k_{AB}) h^{AC} X_C
- \frac{1}{N} (K_{00} - N^AK_{0A})\,n \,, \\
\fl
{\cal D}_A \eta &=& k_{AB}\,h^{BC} X_C
- \frac{1}{N}( K_{0A} - N^B K_{AB}  )\, n\,, \\
\fl
\dot{n} &=& K_{0A}  h^{AB} X_B
- \frac{1}{N} (K_{00} - N^AK_{0A})\,\eta\,, \\
\fl
{\cal D}_A n &=& K_{AB} h^{BC} X_C
- \frac{1}{N}( K_{0A}  - N^B K_{AB}  )\,\eta\,. 
\end{eqnarray}
Where $K_{00}, K_{0A}, K_{AB}$ are the projections of the extrinsic
curvature tensor $K_{ab}$, defined in (\ref{eq:pr1}), (\ref{eq:pr2}),
(\ref{eq:pr3}).

We record also the time derivative of the spatial metric $h_{AB}$
and its determinant,
\begin{eqnarray}
\dot{h}_{AB} &=& 2 {\cal D}_{(A}N_{B)} + 2 N k_{AB}\,, \\
\dot{h} &=& 2h(N h^{AB} k_{AB}  +  {\cal D}_A N^{A})\,.
\end{eqnarray}

\vspace{1cm}

\noindent{\bf APPENDIX B: POISSON CONSTRAINTS ALGEBRA}

\vspace{.5cm}

In this Appendix we collect the Poisson constraints algebra 
first for the DNG model,  and then for the model quadratic in the
mean extrinsic curvature (\ref{eq:haction2}).

Consider first the constraints (\ref{eq:C}), (\ref{eq:CA}) 
for the DNG model.
We define the phase space functions 
\begin{eqnarray}
C_\lambda &=& \int_\Sigma
\lambda \left( p^{2} + \mu^{2} h \right)\,,  \\
V_{\vec{\lambda}} &=& \int_\Sigma
\lambda^{A} p \cdot X_A\,.
\end{eqnarray}
They satisfy the Poisson algebra
\begin{eqnarray}
\left\lbrace 
C_\lambda , C_{\lambda'} 
\right\rbrace  &=&  V_{\vec{\lambda*}}\,,   
\label{eq:a11}\\
\left\lbrace 
V_{\vec{\lambda}} ,C_{\lambda} 
\right\rbrace  &=&   C_{
{\cal L}_{\vec{\lambda}}\lambda}\,,  \\
\left\lbrace 
V_{\vec{\lambda}} , V_{\vec{\lambda'}} 
\right\rbrace  &=&  V_{[\vec{\lambda},\vec{\lambda'}]}\,, 
\end{eqnarray}
where in (\ref{eq:a11}) we have defined $\lambda^{*A} =
4\mu^{2}hh^{AB}(\lambda \partial_B \lambda' -
\lambda' \partial_B \lambda)$, and ${\cal L}_{\vec{\lambda}}$
denotes the spatial Lie derivative along the vector $\lambda^A$.
The constraints (\ref{eq:C}) and
(\ref{eq:CA}) are first class.

Next consider the primary constraints (\ref{eq:c1}), (\ref{eq:c2}),
and
the secondary constraints
(\ref{eq:s1}), (\ref{eq:s2}) 
for the quadratic model
(\ref{eq:haction2}). We define the phase space functions
\begin{eqnarray}
\fl
C_\lambda &=& \int_\Sigma
\lambda P \cdot \dot{X} \,,\nonumber
\\
\fl
V_{\vec{\lambda}} &=& \int_\Sigma
\lambda^{A} P \cdot  X_A\,, \nonumber
\\
\fl
S_\Lambda &=& \int_\Sigma
\Lambda \,{\cal H}_c \nonumber \\
\fl
&=& 
\int_\Sigma \Lambda \left[ p \cdot \dot{X} +
{N^3 \over 2 \sqrt{h}} P^2 
+ 2 N^A P \cdot \partial_A \dot{X} 
+ (N^2 h^{AB} - N^A N^B ) P \cdot {\cal D}_A {\cal D}_B X \right]\,.
\,,
\nonumber
\\
\fl
T_{\vec{\Lambda}} &=& \int_\Sigma
\Lambda^{A}\left( p \cdot X_A 
+ P \cdot \partial_A \dot{X} \right)\,. 
\nonumber
\end{eqnarray}
The Poisson  constraint algebra  of the primary constraints 
is 
\begin{eqnarray}
\{ C_\lambda , C_{\lambda'} \} &=&0\,, \\
\{ C_\lambda , V_{\vec{\lambda}} \} &=&
 V_{\vec{\lambda'}}\,, 
\label{eq:a12} \\
\{ V_{\vec{\lambda}}, 
V_{\vec{\lambda'}} \} &=&0, 
\end{eqnarray}
where $\lambda^{'A}
= \lambda \lambda^{A}$ in (\ref{eq:a12}).

The Poisson constraint algebra between the primary and secondary
constraints, which serves to identify the latter, is
\begin{eqnarray}
\{  C_\lambda ,  S_\Lambda \} &=& -
S_{\lambda \Lambda} - S_{{\cal L}_{\vec{N'}}\lambda} \,, 
\label{eq:pb1}
\\
\left\lbrace C_\lambda , T_{\vec{\Lambda}} \right\rbrace &=& - 
C_{{\cal L}_{\vec{\Lambda}}\lambda}\,, 
\\
\{ V_{\vec{\lambda}} ,   S_\Lambda \} &=&
 - T_{\vec{\Lambda'}} +  C_{{\cal L}_{\vec{\lambda}}\Lambda} - 
V_{\vec{\lambda'}}\,,
\label{eq:pb2}
\\
\left\lbrace V_{\vec{\lambda}} , T_{\vec{\Lambda}} 
\right\rbrace &=&  V_{[{\vec{\lambda}},{\vec{\Lambda}}]}\,, 
\end{eqnarray}
where in (\ref{eq:pb1}) we have defined $N^{'A}=2\Lambda N^{A} $,
in (\ref{eq:pb2})  $\Lambda'^A = \Lambda \lambda^A$, and
$ \lambda'^A = 2 \Lambda N^B {\cal D}_B \lambda^A $.

Finally, the Poisson algebra of the secondary constraints is
\begin{eqnarray}
\left\lbrace S_{\Lambda}, S_{\Lambda'} \right\rbrace &=&
C_{\widetilde{\lambda}}\,,
\label{eq:pb3}
\\
\left\lbrace S_{\Lambda}, T_{\vec{\Lambda}}   \right\rbrace &=&
- S_{{\cal L}_{\vec{\Lambda}}\Lambda} +
V_{\vec{\hat{\lambda}}}\,, 
\label{eq:pb4}
\\
\left\lbrace T_{\vec{\Lambda}} ,
T_{\vec{\Lambda'}}  \right\rbrace &=&
T_{[{\vec{\Lambda}},{\vec{\Lambda'}}]}\,.
\end{eqnarray}
where in (\ref{eq:pb3}) we have defined $
\widetilde{\lambda} = (N^{2} h^{AB} - N^A N^B )
\left( \Lambda {\cal D}_A {\cal D}_B {\Lambda'} -
{\Lambda'} {\cal D}_A {\cal D}_B \lambda \right) $, and in
(\ref{eq:pb4})  $\hat{\lambda}^{A} = \Lambda 
(N^{2} h^{BC} - N^B N^C ) \left( {\cal D}_B {\cal D}_C \Lambda^{A}
+ \Lambda^{D}h^{EA}R_{BECD}   \right) $. $ R_{BECD} $ is the Riemann
tensor on $\Sigma$. 
The constraints are first class. For the special case of 
a relativistic string there is a considerable simplification
in the form of the structure functions of the algebra.

We record also the action of the constraints on
the phase space variables
\begin{eqnarray}
\{ C_\lambda , X \} &=& 0\,, 
\\
\{ C_\lambda , \dot{X} \} &=& - \lambda \dot{X}\,, 
\\
\{ C_\lambda, p \} &=& 0\,, 
\\ 
\{ C_{\lambda}, P \} &=& \lambda P \\
\{ V_{\vec{\lambda}}, X \} &=& 0\,, \\
\{ V_{\vec{\lambda}}, \dot{X} \} &=& - {\cal L}_{\vec{\lambda}} 
X\,,  \\
\{ V_{\vec{\lambda}}, p \} &=& - {\cal L}_{\vec{\lambda}} 
P\,,   
\\
\{ V_{\vec{\lambda}}, P \} &=& 0\,, \\
\{  T_{\vec{\Lambda}}, X \} &=& - {\cal L}_{\vec{\Lambda}} X\,,  \\
\{ T_{\vec{\Lambda}}, \dot{X} \} &=& - {\cal L}_{\vec{\Lambda}} 
\dot{X}\,,  \\
\{ T_{\vec{\Lambda}}, p \} &=& - {\cal L}_{\vec{\Lambda}} p\,,  \\
\{ T_{\vec{\Lambda}}, P \} &=& - {\cal L}_{\vec{\Lambda}} P\,.
\end{eqnarray}
We refrain from writing down the Poisson bracket of the 
secondary constraint $S_\Lambda$ on the phase space variables
because it can be obtained easily from the calculations 
that lead to the Hamilton's equations in Sect. 7.

\end{document}